Version 11
# Do reductionist cures select for holistic diseases? Adaptive chronic infection, structured stress, and medical magic bullets

Rodrick Wallace
The New York State Psychiatric Institute
Robert G. Wallace
Dept. of Biology
City College of New York*

December 31, 2003

## Abstract

With a generalized language-of-thought argument for immune cognition, we model how population-directed structured psychosocial stress can impose an image of itself on the coevolutionary conflict between a highly adaptive chronic infection and the immune response.

We raise the possibility that, for successful 'evolution machines' like HIV and malaria, simplistic individual-oriented magic bullet drug treatments, vaccines, behavior modifications, or other interventions that do not address the fundamental living and working conditions underlying disease ecology will fail to control current epidemics and may select for holistic pathogenic life histories which increase disease virulence.

**Key words**: adaptation, chronic infection, cognition, immune, interpenetration, mutator, phase transition, renormalization, virulence

## Introduction

The first papers in this series (Wallace and Wallace, 2002; Wallace, 2002a), examined culturally-driven variation in HIV transmission and malaria pathology. HIV responds to immune challenge as an evolution machine, generating copious variation and hiding from counterattack in refugia at multiple scales of space, time, and population. *P. falciparum* engages in analogous rapid clonal antigenic variation, and cyto-adherence and sequestration in the deep vasculature, primary mechanisms for escaping from antibody-mediated mechanisms of the host's immune system (e.g. Alred, 1998). Something much like the mutator mechanism, in the sense of Thaler (1999), or 'second order selection' in the sense of Tenallion et al. (2001), appears to generate antigenic variation in the face of immune attack for a large class of pathogens. On the other hand, recent work by DiNoia and Neuberger (2002) outlines the mechanisms by which the immune system's own antibody-producing B-cells engage in a second-order fine tuning of antibody production through an exceedingly high rate of mutation-like transformations, a hypermutation which allows us to respond quickly and effectively to pathogens that we have been exposed to previously (Gearhart, 2002).

Many chronic infections, particularly those which cloak themselves in antigenic 'coats of many colors', are very often marked by distinct 'stages' over the course of disease. For HIV this typically involves an initial viremia triggering an immune response that drives the virus into refugia during an extended asymptomatic period which, with the collapse of the immune system, ends in AIDS. Malaria's most evident 'stages' are expressed as explosive outbursts of rapid parasite replication which facilitate insect-mediated transmission between hosts. HIV, malaria, and a third disease, tuberculosis, account for over five million deaths a year worldwide and exemplify the evolutionary success of multiple-stage chronicity as a life history strategy (Ewald, 2000; Villarreal, et al., 2000).

Here we extend the earlier theoretical analysis of Wallace (2002a), which focused on infection as a sudden 'perturbation'. We will analyze how pathogen life history stages represent a kind of evolutionary punctuation (e.g. Eldredge, 1985) for chronic infection in the face of relentless immune and other selection pressure. For HIV that punctuation may arise from the direct interactions between the virus and the immune system response. In the case of malaria, it may result by means of a 'second order punctuation' through the mutator mechanism (Thaler, 1999) associated with rapid antigenic variation. Elsewhere we study clonal selection in tumorigenesis from such a 'second order' perspective (Wallace et al., 2003).

It is this interpenetration between antagonistic adaptive processes that so defines disease ecosystems. Adami et al. (2000) applied an information theoretic approach to conclude that genomic complexity resulting from evolutionary adaptation can be identified with the amount of information a gene sequence stores about its environment. Lewontin (2000) in essence suggested something of a reverse process, in which environmental complexity represents the amount of information organisms introduce into their environment as a result of their collective actions and interactions. We propose modeling the two 'information sources' provides a more faithful encapsulation of the interactive, multi-scale nature of pathogen-immune

*Address correspondence to Rodrick Wallace, PISCS Inc., 549 W 123 St., Suite 16F, New York, NY, 10027. Telephone (212) 865-4766, email rdwall@ix.netcom.com. Affiliations are for identification only.

dynamics than does the common 'differential paradigm' (e.g., Nowak and May, 2000).

Wallace (2002b) has applied a Rate Distortion argument in the context of imposed renormalization symmetry to obtain evolutionary 'punctuated equilibrium' (Eldredge, 1985) as a consequence of Ademi's mechanism. Here we use the more general Joint Asymptotic Equipartition Theorem (JAEPT) to conclude that pathogenic adaptive response and coupled cognitive immune challenge will be jointly linked in chronic infection, and subject to a transient 'punctuated interpenetration' very similar to evolutionary punctuation. Multiple punctuated transitions, perhaps of mixed 'order', are seen as constituting shifts to the different stages of chronic infection.

Examining paths in parameter space for the renormalization properties of such transitions (i.e., 'universality class tuning' in the sense of Albert and Barabasi, 2001) produces a second order punctuation in the rate at which the selection pressure of the immune system imposes a distorted image of itself onto pathogen structure. This is our version of the mutator, or what Tenallion et al. (2001) call 'second order selection'.

Recognizably similar matters have long been under scrutiny: interactions between the central nervous system (CNS) and the immune system, and between genetic heritage and the immune system have become academically codified through journals with titles such as *Neuroimmunology* and *Immunogenetics*. Elsewhere (Wallace and Wallace, 2002) we introduced another complication by arguing that the culture in which humans are socially embedded also interacts with individual immune systems to form a composite entity that we labeled an *immunocultural condensation* (ICC). It is, we will argue here, the joint entity of immune, CNS, and embedding sociocultural cognition that engages in orders of 'punctuated interpenetration' with an adaptive chronic infectious challenge. Similar arguments are already in the French literature (e.g. Combes, 2000). The interpenetration has fundamental implications for the kinds of individual-level disease interventions largely pursued today.

Before entering the formal thicket, it is important to highlight some general considerations. First, the information theory approach we adopt is notorious for providing 'existence theorems' whose 'representation', to use physics jargon, is arduous. For example, although the Shannon Coding Theorem implied the possibility of highly efficient coding schemes as early as 1949, it took more than forty years for practical 'turbo codes' to be created. The program we propose is unlikely to be any less difficult.

Second, we are invoking information theory variants of the fundamental limit theorems of probability. These are independent of exact mechanisms, but constrain the behavior of those mechanisms. For example, although not all processes involve long sums of independent stochastic variables, those that do, regardless of the individual variable distribution, collectively follow a Normal distribution as a consequence of the Central Limit Theorem. Similarly, the games of chance in a Las Vegas casino are all quite different, but nonetheless the success of 'strategies' for playing them is strongly and systematically constrained by the Martingale Theorem, regardless of game details. We similarly propose that languages-on-networks and languages-that-interact, as a consequence of the limit theorems of information theory, will be subject to regularities of punctuation and 'generalized Onsager relations', regardless of detailed mechanisms, as important as the latter may be.

Finally, just as parametric statistics are imposed, at least as a first approximation, on sometimes questionable experimental situations, relying on the robustness of the Central Limit Theorem to carry us through, we will invoke here a similar heuristic approach for the information theory limit theorems we define.

We begin with a description of cognitive process, including Cohen's (2000) immune cognition, in terms of an information source, a 'language' constrained by the Shannon-McMillan or Asymptotic Equipartition Theorem, and its Rate Distortion or Joint Asymptotic Equipartition and other variants for interacting sources.

### Cognition as language

Adams (2003) reviews in some detail the 'informational turn in philosophy', that is, the relatively recent application of communication theory formalism and concepts to the understanding of cognitive process. One of the first reasonably successful syntheses was that of Dretske (1981, 1988, 1992, 1993, 1994), whose work Adams describes as follows:

> "It is not uncommon to think that information is a commodity generated by things with minds. Let's say that a naturalized account puts matters the other way around, viz. it says that minds are things that come into being by purely natural causal means of exploiting the information in their environments. This is the approach of Dretske as he tried consciously to unite the cognitive sciences around the well-understood mathematical theory of communication..."

Dretske himself (1994) writes:

> "Communication theory can be interpreted as telling one something important about the conditions that are needed for the transmission of information as ordinarily understood, about what it takes for the transmission of semantic information. This has tempted people... to exploit [information theory] in semantic and cognitive studies, and thus in the philosophy of mind.
> ...Unless there is a statistically reliable channel of communication between [a source and a receiver]... no signal can carry semantic information... [thus] the channel over which the [semantic] signal arrives [must satisfy] the appropriate statistical constraints of communication theory."

Here we redirect attention from the informational content or 'meaning' of individual symbols, i.e. the province of semantics, back to the statistical properties of long trains of symbols emitted by an 'information source'. As Dretske so clearly saw, this allows scientific inference on the necessary conditions for cognitive process, including immune cognition.

Atlan and Cohen (1998) and Cohen (2000), following a long tradition (e.g., Grossman, 1989; Tauber, 1998), argue that the essence of immune function is cognitive, involving comparison of a perceived antigenic signal with an internal, learned picture of the world, and then, upon that comparison, the choice of one response from a large repertoire of possible responses. Following the approach of Wallace (2000, 2002a), we make a 'weak', and hence very general, model of that process.

Pattern recognition-and-response, as we characterize it, proceeds by convoluting (i.e. comparing) an incoming external 'sensory' antigenic signal with an internal 'ongoing activity' – the 'learned picture of the world' – and, at some point, triggering an appropriate action based on a decision that the pattern of sensory activity requires a response. We will, fulfilling Atlan and Cohen's (1998) criterion of meaning-from-response, define a language's contextual meaning entirely in terms of system output, leaving the question of how such a pattern recognition system is 'trained' in the first place for future work.

The model here is as follows.

A pattern of sensory (antigenic) input is convoluted (compared) with internal 'ongoing' activity to create a path of convoluted signal $x = (a_0, a_1, ..., a_n, ...)$. This path is fed into a highly nonlinear 'decision oscillator' which generates an output $h(x)$ that is an element of one of two (presumably) disjoint sets $B_0$ and $B_1$. We take

$$B_0 \equiv b_0, ..., b_k,$$

$$B_1 \equiv b_{k+1}, ..., b_m.$$

Thus we permit a graded response, supposing that if

$$h(x) \in B_0$$

the pattern is not recognized, and if

$$h(x) \in B_1$$

the pattern is recognized and some action $b_j, k+1 \leq j \leq m$ takes place.

We are interested in paths $x$ which trigger pattern recognition-and-response exactly once. That is, given a fixed initial state $a_0$, such that $h(a_0) \in B_0$, we examine all possible subsequent paths $x$ beginning with $a_0$ and leading exactly once to the immune event $h(x) \in B_1$. Thus $h(a_0, ..., a_j) \in B_0$ for all $j < m$, but $h(a_0, ..., a_m) \in B_1$.

For each positive integer $n$ let $N(n)$ be the number of paths of length $n$ which begin with some particular $a_0$ having $h(a_0) \in B_0$ and lead to the condition $h(x) \in B_1$. We shall call such paths 'meaningful' and assume $N(n)$ to be considerably less than the number of all possible paths of length $n$ – pattern recognition-and-response is comparatively rare. We further assume that the finite limit

$$H \equiv \lim_{n \to \infty} \frac{\log[N(n)]}{n}$$

both exists and is independent of the path $x$. We will – not surprisingly – call such a pattern recognition-and-response cognitive process *ergodic*.

We may thus define an ergodic information source $\mathbf{X}$ associated with stochastic variates $X_j$ having joint and conditional probabilities $P(a_0, ..., a_n)$ and $P(a_n|a_0, ..., a_{n-1})$ such that appropriate joint and conditional Shannon uncertainties may be defined which satisfy the relations

$$H[\mathbf{X}] = \lim_{n \to \infty} \frac{\log[N(n)]}{n} =$$

$$\lim_{n \to \infty} H(X_n|X_0, ..., X_{n-1}) =$$

$$\lim_{n \to \infty} \frac{H(X_0, ..., X_n)}{n}.$$

The Shannon uncertainties $H(...)$ are defined in terms of cross-sectional sums of the form $-\sum_k P_k \log[P_k]$, where the $P_k$ constitute a probability distribution.

We say this information source is *dual* to the ergodic cognitive process.

Different 'languages' will, of course, be defined by different divisions of the total universe of possible responses into different pairs of sets $B_0$ and $B_1$, or by requiring more than one response in $B_1$ along a path. However, like the use of different distortion measures in the Rate Distortion Theorem (e.g. Cover and Thomas, 1991), it seems obvious that the underlying dynamics will all be qualitatively similar.

Meaningful paths – creating an inherent grammar and syntax – are defined entirely in terms of system response, as Atlan and Cohen (1998) propose. See Wallace (2002a) for explicit application of this formalism to the stochastic neuron.

We will eventually parametize the information source uncertainty of this dual information source with respect to one or more variates, writing, e.g. $H[\mathbf{K}]$, where $\mathbf{K} \equiv (K_1, ..., K_s)$ represents a vector in a parameter space. Let the vector $\mathbf{K}$ follow some path in time, i.e. trace out a generalized line or surface $\mathbf{K}(t)$. We will, following the argument of Wallace (2002b), assume that the probabilities defining $H$, for the most part, closely track changes in $\mathbf{K}(t)$, so that along a particular 'piece' of a path in parameter space the information source remains as close to memoryless and ergodic as is needed for the mathematics to work. Between pieces we impose phase transition characterized by a renormalization symmetry, in the sense of Wilson (1971).

We will call such an information source 'piecewise memoryless ergodic'.

Iterating the argument on paths of 'tuned' sets of renormalization parameters gives a second order punctuation in the rate at which primary interacting information sources come to match each other in a distorted manner, the essence of adaptation or interpenetration.

### Introduction to the general argument

Taking a formal Dretske-style language-of-thought description of immune cognition as a starting point, Wallace (2002a) has explored host response to sudden pathogenic challenge, using a mathematical model of the generalized 'cognitive condensation' that characterizes human biology. Suppose the pathogen avoids extirpation by that response, but, changing its coat or hiding within refugia, becomes an established invading population. While the immune system is cognitive, the pathogen is adaptive.

We suppose that the host's generalized CNS and immuno-cultural condensation can be represented by a sequence of 'states', the 'path' $x \equiv x_0, x_1, ...$. Similarly, we assume the pathogen population can be represented by the path $y \equiv y_0, y_1, ...$. These paths are, however, both very highly structured and serially correlated and can, in fact, be represented by 'information sources' **X** and **Y**. Since the host and parasite population interact, these sequences of states are not independent, but are jointly serially correlated. We can, then, define a path of sequential pairs as $z \equiv (x_0, y_0), (x_1, y_1), ...$. The essential content of the Joint Asymptotic Equipartition Theorem (JAEPT), one of the fundamental limit theorems of 20th Century mathematics, is that the set of joint paths $z$ can be partitioned into a relatively small set of high probability termed *jointly typical*, and a much larger set of vanishingly small probability. Further, according to the JAEPT, the *splitting criterion* between high and low probability sets of pairs is the mutual information

$$I(X,Y) = H(X) - H(X|Y) = H(X) + H(Y - H(X,Y)$$

where $H(X), H(Y), H(X|Y)$ and $H(X,Y)$ are, respectively, the Shannon uncertainties of $X$ and $Y$, their conditional uncertainty, and their joint uncertainty. See Cover and Thomas (1991) for mathematical details. Similar approaches to neural process have been recently adopted by Dimitrov and Miller (2001).

The high probability pairs of paths are, in this formulation, all equiprobable, and if $N(n)$ is the number of jointly typical pairs of length $n$, then

$$I(X,Y) = \lim_{n \to \infty} \frac{\log[N(n)]}{n}.$$

Generalizing the earlier language-on-a-network models of Wallace and Wallace (1998, 1999), we suppose there is a 'chronic coupling parameter' $P$ representing the degree of linkage between host's ICC/CNS condensation and the parasite population, and set $K = 1/P$, following the development of those earlier studies. Then we have

$$I[K] = \lim_{n \to \infty} \frac{\log[N(K,n)]}{n}.$$

The essential 'homology' between information theory and statistical mechanics lies in the similarity of this expression with the infinite volume limit of the free energy density. If $Z(K)$ is the statistical mechanics partition function derived from the system's Hamiltonian, then the free energy density is determined by the relation

$$F[K] = \lim_{V \to \infty} \frac{\log[Z(K)]}{V}.$$

$F$ is the free energy density, $V$ the system volume and $K = 1/T$, where $T$ is the system temperature.

We and others argue at some length (Wallace and Wallace, 1998, 1999; Rojdestvensky and Cottam, 2000) that this is indeed a systematic mathematical homology which, we contend, permits importation of renormalization symmetry into information theory. Imposition of invariance under renormalization on the mutual information splitting criterion $I(X,Y)$ implies the existence of phase transitions analogous to learning plateaus or punctuated evolutionary equilibria in the relations between host and pathogen. An extensive mathematical development will be presented in the next section.

The physiological details of mechanism, we speculate, will be particularly captured by the definitions of coupling parameter, renormalization symmetry, and, perhaps, the distribution of the renormalization across agency, a matter we treat below.

Here, however, these changes are perhaps better described as 'punctuated interpenetration' between the challenged cognitive condensation of the host and the adaptive abilities of the pathogen.

Even more elaborate developments are possible. For example, in the next section we explore canonical patterns of transition between disease stages that emerge quite naturally. We reiterate that the details are highly dependent on the choice of renormalization symmetry, which is likely to reflect details of mechanism – the manner in which the dynamics of the forest are dependent on the detailed physiology of trees, albeit in a many-to-one manner. Renormalization properties are not likely to follow simple physical analogs, and may well be subject to characteristic distributions. The algebra is straightforward if complicated, and given later. Following Nesbitt et al. (2001), however, any 'cognitive' process is likely to show significant cultural variation, and even distribution of properties.

**Representations of the general argument**

**1. Language-on-a-network models.** Earlier work in this series addressed the problem of how a 'language', in a large sense, 'spoken' on a network structure responds as properties of the network change. The language might be spoken, pattern recognition, or cognition. The network might be social, chemical, or neural. The properties of interest were the magnitude of 'strong' or 'weak' ties which, respectively, either disjointly partitioned the network or linked it. These would be analogous to local and mean-field couplings in physical systems.

We fix the magnitude of strong ties, but vary the index of weak ties between components, which we call $P$, taking $K = 1/P$. For neural networks $P$ is just proportional to the number of training cycles, suggesting that, for interacting cognitive/adaptive systems, $P$ may be proportional to the number of 'challenge cycles', likely indexed by human diurnal or other activity patterns, or perhaps even those of the parasite itself.

We assume the piecewise, adiabatically memoryless ergodic information source (Wallace, 2002b) depends on three parameters, two explicit and one implicit. The explicit are $K$ as above and an 'external field strength' analog $J$, which gives a 'direction' to the system. We will, in the limit, set $J = 0$.

The implicit parameter, which we call $r$, is an inherent generalized 'length' characteristic of the phenomenon, on which $J$ and $K$ are defined. That is, we can write $J$ and $K$ as functions of averages of the parameter $r$, which may be quite complex, having nothing at all to do with conventional ideas of space: For example $r$ may be defined by the degree of niche partitioning in ecosystems or separation in social structures.

For a given generalized language of interest with a well defined (piecewise adiabatically memoryless) ergodic source uncertainty $H$ we write

$$H[K, J, \mathbf{X}]$$

Imposition of invariance of $H$ under a renormalization transform in the implicit parameter $r$ leads to expectation of both a critical point in $K$, which we call $K_C$, reflecting a phase transition to or from collective behavior across the entire array, and of power laws for system behavior near $K_C$. Addition of other parameters to the system, e.g. some $V$, results in a 'critical line' or surface $K_C(V)$.

Let $\kappa = (K_C - K)/K_C$ and take $\chi$ as the 'correlation length' defining the average domain in $r$-space for which the information source is primarily dominated by 'strong' ties. We begin by averaging across $r$-space in terms of 'clumps' of length $R$. Then, taking Wilson's (1971) analysis as a starting point, we choose the renormalization relations as

$$H[K_R, J_R, \mathbf{X}] = f(R) H[K, J, \mathbf{X}]$$

$$\chi(K_R, J_R) = \frac{\chi(K, J)}{R},$$

(1)

with $f(1) = 1$ and $J_1 = J, K_1 = K$. The first of these equations significantly extends Wilson's treatment. It states that 'processing capacity,' as indexed by the source uncertainty of the system, representing the 'richness' of the generalized language, grows monotonically as $f(R)$, which must itself be a dimensionless function in $R$, since both $H[K_R, J_R]$ and $H[K, J]$ are themselves dimensionless. Most simply, this would require that we replace $R$ by $R/R_0$, where $R_0$ is the 'characteristic length' for the system over which renormalization procedures are reasonable, then set $R_0 \equiv 1$, i.e. measure length in units of $R_0$. Wilson's original analysis focused on free energy density. Under 'clumping', densities must remain the same, so that if $F[K_R, J_R]$ is the free energy of the clumped system, and $F[K, J]$ is the free energy density before clumping, then Wilson's equation (4) is $F[K, J] = R^{-3} F[K_R, J_R]$, i.e.

$$F[K_R, J_R] = R^3 F[K, J].$$

Remarkably, the renormalization equations are solvable for a broad class of functions $f(R)$, or more precisely, $f(R/R_0), R_0 \equiv 1$.

The second relation just states that the correlation length simply scales as $R$.

Other, very subtle, symmetry relations – not necessarily based on the elementary physical analog we use here – may well be possible. For example McCauley, (1993, p.168) describes the highly counterintuitive renormalization relations needed to understand phase transition in simple 'chaotic' systems. This is an important subject for future research, since we suspect that biological or social systems may alter their renormalization properties in response to external pressures.

To begin, following Wilson, we take $f(R) = R^d$ for some real number $d > 0$, and restrict $K$ to near the 'critical value' $K_C$. If $J \to 0$, a simple series expansion and some clever algebra (Wilson, 1971; Binney et al., 1986) gives

$$H = H_0 \kappa^\alpha$$

$$\chi = \frac{\chi_0}{\kappa^s}$$

(2)

where $\alpha, s$ are positive constants. We provide more biologically relevant examples below.

Further from the critical point matters are more complicated, appearing to involve 'Generalized Onsager Relations' and a kind of thermodynamics associated with a Legendre transform (Wallace, 2002a).

An essential insight is that *regardless of the particular renormalization properties, sudden critical point transition is possible in the opposite direction for this model*. That is, we go from a number of independent, isolated and fragmented systems operating individually and more or less at random, into a single large, interlocked, coherent structure, once the parameter $K$, the inverse strength of weak ties, falls below threshold, or, conversely, once the strength of weak ties parameter $P = 1/K$ becomes large enough.

Thus, increasing nondisjunctive weak ties between them can bind several different 'language' processes into a single, embedding hierarchical metalanguage which contains each as a linked subdialect.

To reiterate somewhat, this heuristic insight can be made more exact using a rate distortion argument (or, more generally, using the Joint Asymptotic Equipartition Theorem) as follows (Wallace, 2002a, b):

Suppose that two ergodic information sources $\mathbf{Y}$ and $\mathbf{B}$ begin to interact, to 'talk' to each other, i.e. to influence each other in some way so that it is possible, for example, to look at the output of $\mathbf{B}$ – strings $b$ – and infer something about the behavior of $\mathbf{Y}$ from it – strings $y$. We suppose it possible to define a retranslation from the B-language into the Y-language through a deterministic code book, and call $\hat{\mathbf{Y}}$ the translated information source, as mirrored by $\mathbf{B}$.

Define some distortion measure comparing paths $y$ to paths $\hat{y}, d(y, \hat{y})$ (Cover and Thomas, 1991). We invoke the Rate Distortion Theorem's mutual information $I(Y, \hat{Y})$, which is the splitting criterion between high and low probability pairs of paths. Impose, now, a parametization by an inverse coupling strength $K$, and a renormalization symmetry representing the global structure of the system coupling. This may be much different from the renormalization behavior of the individual components. If $K < K_C$, where $K_C$ is a critical point (or surface), the two information sources will be closely coupled enough to be characterized as condensed.

In the absence of a distortion measure, we can invoke the Joint Asymptotic Equipartition Theorem to obtain a similar result.

We suggest in particular that detailed biochemical and molecular coupling mechanisms will be sharply constrained through regularities of grammar and syntax imposed by limit theorems associated with phase transition.

Wallace and Wallace (1998, 1999) use this approach to address speciation, coevolution and group selection in a relatively unified fashion. These papers, and those of Wallace and Fullilove (1999) and Wallace (2002a), further describe how biological or social systems might respond to gradients in information source uncertainty and related quantities when the system is away from phase transition. Language-on-network systems, as opposed to physical systems, appear to diffuse away from concentrations of an 'instability' construct which is related to a Legendre transform of information source uncertainty, in much the same way entropy is the Legendre transform of free energy density in a physical system. The parametized 'instability', $Q[K]$, is defined from the principal splitting criterion by the relation

$$Q[K] = -KdH[K]/dK$$

$$Q[K] = -KdI[K]/dK$$

(3)

where $H[K]$ and $I[K]$ are, respectively, information source uncertainty or mutual information in the Asymptotic Equipartition, Rate Distortion, or Joint Asymptotic Equipartition Theorems.

**2. 'Biological' phase transitions.** Equation (2) states that the information source and the correlation length, the degree of coherence on the underlying network, scale under renormalization clustering in chunks of size $R$ as

$$H[K_R, J_R]/f(R) = H[J, K]$$

$$\chi[K_R, J_R]R = \chi(K, J),$$

with $f(1) = 1, K_1 = K, J_1 = J$, where we have slightly rearranged terms.

Differentiating these two equations with respect to $R$, so that the right hand sides are zero, and solving for $dK_R/dR$ and $dJ_R/dR$ gives, after some consolidation, expressions of the form

$$dK_R/dR = u_1 d\log(f)/dR + u_2/R$$

$$dJ_R/dR = v_1 J_R d\log(f)/dR + \frac{v_2}{R} J_R.$$

(4)

The $u_i, v_i, i = 1, 2$ are functions of $K_R, J_R$, but not explicitly of $R$ itself.

We expand these equations about the critical value $K_R = K_C$ and about $J_R = 0$, obtaining

$$dK_R/dR = (K_R - K_C)yd\log(f)/dR + (K_R - K_C)z/R$$

$$dJ_R/dR = wJ_R d\log(f)/dR + xJ_R/R.$$

(5)

The terms $y = du_1/dK_R|_{K_R=K_C}, z = du_2/dK_R|_{K_R=K_C}, w = v_1(K_C, 0), x = v_2(K_C, 0)$ are constants.

Solving the first of these equations gives

$$K_R = K_C + (K - K_C)R^z f(R)^y,$$

(6)

again remembering that $K_1 = K, J_1 = J, f(1) = 1$.

Wilson's essential trick is to iterate on this relation, which is supposed to converge rapidly (Binney, 1986), assuming that for $K_R$ near $K_C$, we have

$$K_C/2 \approx K_C + (K - K_C)R^z f(R)^y.$$

(7)

We iterate in two steps, first solving this for $f(R)$ in terms of known values, and then solving for $R$, finding a value $R_C$ that we then substitute into the first of equations (1) to obtain an expression for $H[K, 0]$ in terms of known functions and parameter values.

The first step gives the general result

$$f(R_C) \approx \frac{[KC/(KC - K)]^{1/y}}{2^{1/y} R_C^{z/y}}.$$

(8)

Solving this for $R_C$ and substituting into the first of equation (1) gives, as a first iteration of a far more general procedure (e.g. Shirkov and Kovalev, 2001)

$$H[K,0] \approx \frac{H[K_C/2, 0]}{f(R_C)} = \frac{H_0}{f(R_C)}$$

$$\chi(K,0) \approx \chi(K_C/2, 0)R_C = \chi_0 R_C \tag{9}$$

which are the essential relationships.

Note that a power law of the form $f(R) = R^m, m = 3$, which is the direct physical analog, may not be biologically reasonable, since it says that 'language richness' can grow very rapidly as a function of increased network size. Such rapid growth is simply not observed.

If we take the biologically realistic example of non-integral 'fractal' exponential growth,

$$f(R) = R^\delta, \tag{10}$$

where $\delta > 0$ is a real number which may be quite small, we can solve equation (8) for $R_C$, obtaining

$$R_C = \frac{[KC/(KC-K)]^{[1/(\delta y + z)]}}{2^{1/(\delta y + z)}} \tag{11}$$

for $K$ near $K_C$. Note that, for a given value of $y$, we might want to characterize the relation $\alpha \equiv \delta y + z =$ constant as a "tunable universality class relation" in the sense of Albert and Barabasi (2002).

Substituting this value for $R_C$ back into equation (9) gives a somewhat more complex expression for $H$ than equation (2), having three parameters, i.e. $\delta, y, z$.

A more biologically interesting choice for $f(R)$ is a logarithmic curve that 'tops out', for example

$$f(R) = m \log(R) + 1. \tag{12}$$

Again $f(1) = 1$.

Using Mathematica 4.2 to solve equation (8) for $R_C$ gives

$$R_C = [\frac{Q}{LambertW[Q \exp(z/my)]}]^{y/z}, \tag{13}$$

where

$$Q \equiv [(z/my)2^{-1/y}[KCKC - K]]^{1/y}.$$

The transcendental function LambertW(x) is defined by the relation

$$LambertW(x) \exp(LambertW(x)) = x.$$

It arises in the theory of random networks and in renormalization strategies for quantum field theories.

An asymptotic relation for $f(R)$ would be of particular biological interest, implying that 'language richness' increases to a limiting value with population growth. Such a pattern is broadly consistent with calculations of the degree of allelic heterozygosity as a function of population size under a balance between genetic drift and neutral mutation (Hartl and Clark, 1997; Ridley, 1996). Taking

$$f(R) = \exp[m(R-1)/R] \tag{14}$$

gives a system which begins at 1 when R=1, and approaches the asymptotic limit $\exp(m)$ as $R \to \infty$. Mathematica 4.2 finds

$$R_C = \frac{my/z}{LambertW[S]} \tag{15}$$

where

$$S \equiv (my/z) \exp(my/z)[2^{1/y}[KC/(KC-K)]^{-1/y}]^{y/z}.$$

(15)

These developments indicate the possibility of taking the theory significantly beyond arguments by abduction from simple physical models, although the notorious difficulty of implementing information theory existence arguments will undoubtedly persist.

**3. Universality class distribution.** Physical systems undergoing phase transition usually have relatively 'pure' renormalization properties, with quite different systems clumped into the same 'universality class', having fixed exponents at transition (e.g. Binney, 1986). Biological and social phenomena may be far more complicated:

If we suppose the system of interest to be a mix of subgroups with different values of some significant renormalization parameter $m$ in the expression for $f(R, m)$, according to a distribution $\rho(m)$, then we expect the first expression in equation (1) to generalize as

$$H[K_R, J_R] = <f(R,m)> H[K, J]$$
$$\equiv H[K, J] \int f(R,m)\rho(m)dm.$$

(16)

If $f(R) = 1 + m\log(R)$ then, given any distribution for $m$, we simply obtain

$$<f(R)> = 1 + <m>\log(R)$$

(17)

where $<m>$ is simply the mean of $m$ over that distribution.

Other forms of $f(R)$ having more complicated dependencies on the distributed parameter or parameters, like the power law $R^\delta$, do not produce such a simple result. Taking $\rho(\delta)$ as a normal distribution, for example, gives

$$<R^\delta> = R^{<\delta>} \exp[(1/2)(\log(R^\sigma))^2],$$

(18)

where $\sigma^2$ is the distribution variance. The renormalization properties of this function can be determined from equation (8), and is left to the reader as an exercise, best done in Mathematica 4.2.

Thus the information dynamic phase transition properties of mixed systems will not in general be simply related to those of a single subcomponent, a matter of possible empirical importance: If sets of relevant parameters defining renormalization 'universality classes' are indeed distributed, experiments observing 'pure' phase changes may be very difficult. Tuning among different possible renormalization strategies in response to external pressures would result in even greater ambiguity in recognizing and classifying information dynamic phase transitions.

We believe that important aspects of mechanism may be reflected in the combination of renormalization properties and the details of their distribution across subsystems.

In sum, real biological, social, or 'biopsychosocial' systems are likely to have very rich patterns of phase transition which may not display the simplistic, indeed, literally elemental, purity familiar to physicists. Overall mechanisms will, we believe, still remain significantly constrained by our theory, in the general sense of probability limit theorems.

**4. Universality class tuning** Next we iterate the general argument onto the process of phase transition itself, obtaining Tenallion's 'second order selection', i.e. the mutator, in a 'natural' manner.

We suppose that a structured environment, which we take itself to be an appropriately regular information source $\mathbf{Y}$ – e.g. the immune system, or more generally, for humans the immunocultural condensation (ICC) – 'engages' a modifiable system – e.g. a pathogen – through selection pressure. The ICC begins to write itself on the pathogen's genetic sequences or protein residues in a distorted manner permitting definition of a mutual information $I[K]$ splitting criterion according to the Rate Distortion or Joint Asymptotic Equipartition Theorems. $K$ is an inverse coupling parameter between system and environment (Wallace, 2002a, b). According to our development, at punctuation – near some critical point $K_C$ – the systems begin to interact very strongly indeed, and we may write, near $K_C$, taking as the starting point the simple physical model of equation (2),

$$I[K] \approx I_0 [\frac{K_C - K}{K_C}]^\alpha.$$

For a physical system $\alpha$ is fixed, determined by the underlying 'universality class'. Here we will allow $\alpha$ to vary, and, in the section below, to itself respond explicitly to selection pressure.

Normalizing $K_C$ and $I_0$ to 1, we obtain,

$$I[K] \approx (1 - K)^\alpha.$$

(19)

The horizontal line $I[K] = 1$ corresponds to $\alpha = 0$, while $\alpha = 1$ gives a declining straight line with unit slope which passes through 0 at $K = 1$. Consideration shows there are progressively sharper transitions between the necessary zero value at $K = 1$ and the values defined by this relation for $0 < K, \alpha < 1$. The rapidly rising slope of transition with declining $\alpha$ is, we assert, of considerable significance.

The instability associated with the splitting criterion $I[K]$ is defined by

$$Q[K] \equiv -K dI[K]/dK = \alpha K (1-K)^{\alpha-1},$$

(20)

and is singular at $K = K_C = 1$ for $0 < \alpha < 1$. Following earlier work (Wallace and Wallace, 1998, 1999; Wallace and Fullilove, 1999; Wallace, 2002a), we interpret this to mean that values of $0 < \alpha \ll 1$ are highly unlikely for real systems, since $Q[K]$, in this model, represents a kind of barrier for information systems.

On the other hand, smaller values of $\alpha$ mean that the system is far more efficient at responding to the adaptive demands imposed by the embedding structured ecosystem, since the mutual information which tracks the matching of internal response to external demands, $I[K]$, rises more and more quickly toward the maximum for smaller and smaller $\alpha$ as the inverse coupling parameter $K$ declines below $K_C = 1$. That is, *systems able to attain smaller $\alpha$ are more adaptive than those characterized by larger values*, in this model, but smaller values will be hard to reach, and can probably be done so only at some considerable physiological or other cost.

The more biologically realistic renormalization strategies given above produce sets of several parameters defining the 'universality class', whose tuning gives behavior much like that of $\alpha$ in this simple example.

We can formally iterate the phase transition argument on this calculation to obtain our version of the mutator, focusing on 'paths' of universality classes.

### The adaptive mutator

Suppose the renormalization properties of a biological or social language-on-a network system at some 'time' $k$ are characterized by a set of parameters $A_k \equiv \alpha_1^k, ..., \alpha_m^k$. Fixed parameter values define a particular universality class for the renormalization. We suppose that, over a sequence of 'times', the universality class properties can be characterized by a path $x_n = A_0, A_1, ..., A_{n-1}$ having significant serial correlations which, in fact, permit definition of an adiabatically piecewise memoryless ergodic information source associated with the paths $x_n$. We call that source **X**.

We further suppose, in the manner of Wallace (2002a, b), that external selection pressure is also highly structured – e.g. the cognitive immune system or, in humans, the ICC – and forms another information source **Y** which interacts not only with the system of interest globally, but specifically with its universality class properties as characterized by **X**. **Y** is necessarily associated with a set of paths $y_n$.

We pair the two sets of paths into a joint path, $z_n \equiv (x_n, y_y)$ and invoke an inverse coupling parameter, $K$, between the information sources and their paths. This leads, by the arguments above, to phase transition punctuation of $I[K]$, the mutual information between **X** and **Y**, under either the Joint Asymptotic Equipartition Theorem or under limitation by a distortion measure, through the Rate Distortion Theorem (Cover and Thomas, 1991). Again, see Wallace (2002a, b) for more details of the argument. The essential point is that $I[K]$ is a splitting criterion under these theorems, and thus partakes of the homology with free energy density which we have invoked above.

Activation of universality class tuning, our version of the mutator, then becomes itself a punctuated event in response to increasing linkage between organism (i.e., the pathogen) and externally imposed selection or other pressure (i.e., responses of the ICC). Mutation rates become a function of the relationship between the ICC and the pathogen, above and beyond environmental insult alone.

Thaler (1999) has suggested that the mutagenic effects associated with a cell sensing its environment and history could be as exquisitely regulated as transcription. Our invocation of the Rate Distortion or Joint Asymptotic Equipartition Theorems in address of the mutator necessarily means that variation comes to significantly reflect the grammar, syntax, and higher order structures of the embedding processes. This involves far more than a simple 'colored noise' – stochastic excursions about a deterministic 'spine' – and most certainly implies the need for exquisite regulation. Our information theory argument here converges with Thaler's speculation.

In the same paper Thaler further argues that the immune system provides an example of a biological system which ignores conceptual boundaries that separate development from evolution. While evolutionary phenomena are not cognitive in the sense of the immune system (Cohen, 2000), they may still partake of a significant interaction with development, in which the very reproductive mechanisms of a cell, organism, or organization become closely coupled with structured external selection pressure in a manner recognizably analogous to 'ordinary' punctuated evolution.

That is, we argue the staged nature of chronic infectious diseases like HIV and malaria represents a punctuated version of biological interpenetration, in the sense of Lewontin (2000), between a cognitive 'immunocultural condensation' and a highly adaptive pathogen. We further suggest that this punctuated interpenetration may have both first–i.e., direct– and second order characteristics, involving cross interactions between direct cognitive effects of the immune system or immunocultural condensation, or, more generally, of the ICC and the mutator mechanisms of both the immune system and its pathogen targets.

### Population-directed structured stress and the pathogenic coat of many colors

As we discuss elsewhere (Wallace and Wallace, 2002; Wallace, 2002a), structured psychosocial stress directed at populations, by policy choice or as unforeseen consequence, constitutes a determining context for immune cognition or, more generally, the immunocultural condensation. We wish to analyze the way structured stress affects the interaction between the cognitive ICC and an adaptive mutator, the principal line of defense against the ICC for a large class of highly successful pathogens. To do this we must extend our theory to three interacting information sources.

The Rate Distortion and Joint Asymptotic Equipartition Theorems are generalizations of the Shannon-McMillan Theorem which examine the interaction of two information sources, with and without the constraint of a fixed average distortion. We conduct one more iteration, and require a generalization of the SMT in terms of the splitting criterion for triplets as opposed to single or double stranded patterns. The tool for this is at the core of what is termed *network information theory* [Cover and Thomas, 1991, Theorem 14.2.3]. Suppose we have three (piecewise adiabatically memoryless) ergodic information sources, $Y_1, Y_2$ and $Y_3$. We assume $Y_3$ constitutes a critical embedding context for $Y_1$ and $Y_2$ so that, given three sequences of length $n$, the probability of a particular triplet of sequences is determined by *conditional probabilities with respect to* $Y_3$:

$$P(Y_1 = y_1, Y_2 = y_2, Y_3 = y_3) =$$

$$\Pi_{i=1}^n p(y_{1i}|y_{3i}) p(y_{2i}|y_{3i}) p(y_{3i}).$$

(21)

That is, $Y_1$ and $Y_2$ are, in some measure, driven by their interaction with $Y_3$

Then, in analogy with previous analyses, triplets of sequences can be divided by a splitting criterion into two sets, having high and low probabilities respectively. For large $n$ the number of triplet sequences in the high probability set will be determined by the relation [Cover and Thomas, 1992, p. 387]

$$N(n) \propto \exp[nI(Y_1; Y_2|Y_3)],$$

(22)

where splitting criterion is given by

$$I(Y_1; Y_2|Y_3) \equiv$$

$$H(Y_3) + H(Y_1|Y_3) + H(Y_2|Y_3) - H(Y_1, Y_2, Y_3)$$

We can then examine mixed cognitive/adaptive phase transitions analogous to learning plateaus (Wallace, 2002b) in the splitting criterion $I(Y_1, Y_2|Y_3)$, which characterizes the synergistic interaction between structured psychosocial stress, the ICC, and the pathogen's adaptive mutator. These transitions delineate the various stages of the chronic infection, which are embodied in the slowly varying 'piecewise adiabatically memoryless ergodic' phase between transitions. Again, our results are exactly analogous to the Eldredge/Gould model of evolutionary punctuated equilibrium.

We can, if necessary, extend this model to any number of interacting information sources, $Y_1, Y_2, ..., Y_s$ conditional on an external context $Z$ in terms of a splitting criterion defined by

$$I(Y_1; ...; Y_s|Z) = H(Z) + \sum_{j=1}^s H(Y_j|Z) - H(Y_1, ..., Y_s, Z),$$

(23)

where the conditional Shannon uncertainties $H(Y_j|Z)$ are determined by the appropriate direct and conditional probabilities.

While this argument has been focused on complex parasites like malaria which may have mutator mechanisms determining behavior of their antigenic coat of many colors, a simplified analysis can be applied directly to HIV, which, as a kind of evolution machine, seems to engage in endless, rapid, direct mutation, and, at broader temporal scales, recombination.

**Discussion and conclusions**

Scientific enterprise encompasses the interaction of facts, tools, and theories, all embedded in a path-dependent political economy which seems as natural to us as air to a bird. Molecular biology, Central Limit Theorem statistics, and 19th century mathematics, presently provide the reductionist tool kit most popular in the study of immune function and disease process. Many essential matters related to the encompassing social, economic, and cultural matrix so fundamental to human biology are simply blindsided, and one is reminded, not very originally, of the joke about the drunk looking for his car keys under a street lamp, "because the light here is better".

The asymptotic limit theorems of probability beyond the Central Limit Theorem, in concert with related formalism adapted from statistical physics, would seem to provide new tools which can generate theoretical speculations of value in obtaining and interpreting empirical results about infection and immune process, particularly regarding the way in which culture is, for human populations, "as much a part of human biology as the enamel on our teeth" (Richerson and Boyd, 1995).

We have, as yet, explored relatively few possibilities: While we can model the interaction of first and second order phenomena in the context of structured stress using network information theory, it is difficult to envision interaction between second order 'tuning' processes, or the mechanics of even higher order effects: can we continue to 'tune the tuners' in a kind of idiotypic hall of mirrors? The mathematics would be straightforward, but the corresponding molecular biology would have to be subtle indeed. While unlikely in general, higher order interpenetration – mutating the mutator – may be observable in certain isolated circumstances, for example the interplay between B-cell hypermutation and a pathogen's hypervariable membrane proteins.

Our model, then, explicitly invokes the possibility of synergistic interaction between the selection pressure of the ICC and the variable antigenic coat of the established invading pathogen population, particularly in the context of embedding patterns of structured psychosocial stress which, to take a Rate Distortion perspective, can literally write an image of itself onto that interaction. The ICC, through its immune hypermutation, may engage in its own second order selection. Hence first, second, and possibly mixed, order interpenetration, in which ICC and pathogen each constitute both selection pressure and selected structure, an interaction which may become a distorted image of enfolding patterns of socioeconomically, historically, and politically determined psychosocial stress.

Human chronic infection cannot, in particular, be simply abstracted as a matter of conflict between the pathogen and the immune system alone. Indeed, the concept of an immune system 'alone' has no meaning within our model, in stark contrast with, for example, the well-stirred Erlenmeyer flask predator-prey population dynamics of Nowak and May (2000).

Our analysis suggests that 'mind-body interaction', culture, and, most particularly, structured psychosocial stress, profoundly influence the course of chronic infection at the individual level. The inherently cognitive nature of human biology, especially the intimate role of culture, would seem to limit the utility of animal models in the study of chronic disease much beyond what is already commonly realized.

Individual and collective history, socioeconomic structure, psychosocial stress and the resulting emotional states, may not be mere adjuncts to what is termed 'basic science' in the medical journals. Rather, they may be as much a part of basic human biology as T-cells. 'Magic bullet' vaccines, therapeutic drugs, or highly-focused medicalized 'social' interventions against HIV disease and other mutagenic parasites – approaches that inherently cannot reckon with socioeconomic, historical, and cultural determinants of health and illness – will likely largely fail as they are overwhelmed by relentless pathogen adaptation and cross-population variation in immune cognition. For chronic infections like HIV and malaria, individual level or limited 'social network' intervention strategies which neglect larger embedding context, and the history of that context, embody a grossly unreal paradigm of basic human biology.

We know that some social systems have succeeded in controlling malaria through, for example, persistent and highly organized programs of insect vector control. For HIV, of course, humans are both the intermediate vector and primary host. The larger social context, then, plays a fundamental role in the individual- and population-level decisions that promote or decelerate the HIV epidemic (Wallace and Wallace 1998, Schoepf et al. 2000). Recent work by one of us (R.G. Wallace, in press) suggests that, alone, individual-level antiretroviral treatment of the HIV epidemic may constitute a selection pressure forcing evolutionary changes in HIV life history, including, in one possibility, a more rapid onset of AIDS. A key result is that increasing infection survivorship and decreasing the transition rate from the asymptomatic stage to AIDS, as drug regimens aim to do, may induce the greatest increase in infection population growth. Because infection survivorship is physiologically enmeshed with host survivorship the asymptomatic stage becomes under the drug regimens a demographic shield against epidemiological intervention. In other words, HIV may use processes at one level of biological organization to defend itself against cures directed at it at other levels. Any successful intervention, then, must display a comparable multidimensionality.

Levins and Lewontin (1985; Levins 1998) have pointed out that reductionist approaches to studying biological phenomena, including disease, are the outgrowth of social decisions about the role and nature of science, as well as about what variables to include in models and experimental tests. Internalizing causality by assuming the whole is the sum of its parts, as reductionism does, is only one of several valid methodologies. Other approaches (Wallace and Wallace 1998, Oyama et al. 2001) permit the study of diseases, and the synergistic interaction of their causes, across scales of space, time, and departmental discipline.

We take a step further here. Our work, here and cited, suggests that study itself can affect organisms, in the population biology equivalent of Heisenberg's uncertainty principle. The act of observation – and subsequent interventions on which they are based – change the phenomenon observed. It is apparent the reductionist approach now *selects for* holistic or dialectical pathogens. Reductionism's wholesale application, while succeeding against diseases such as polio and smallpox, a welcome development notwithstanding, now selects for diseases that are characterized by complex sociogeographies, multiple hosts, and multidimensional interactions across scale. We've discussed here the HIV, malaria, and tuberculosis epidemics that affect so many worldwide. In industrial countries, heart disease, cancer, and obesity take their toll; so-called diseases of affluence the poorest and most marginalized typically suffer the worst (Wallace et al. 2003; Wallace et al., in press).

In ecology we learn that sources of mortality compete. While pharmaceuticals, surgery, and individual-level risk reduction interventions quash what we might call reductionist sources of mortality, both within individuals and populations, the pathogenic playing field appears now tilted towards holistic diseases we are unable to address because of the restricted scientific practice pursued.

Our model, by contrast, raises the possibility of effective 'integrated pathogen management' (IPM) programs through synergistic combinations of social, ecological, and medical interventions. IPM far transcends 'medical' strategies that amount to little more than a kind of pesticide application, an approach currently being abandoned in agriculture as simply inadequate to address pathogen evolutionary strategies.

### Acknowledgments

This research was partially supported through NIEHS Grant I-P50-ES09600-05, and benefited from earlier monies given Rodrick Wallace under an Investigator Award in Health Policy Research from the Robert Wood Johnson Foundation.

### References

Adami C., C. Ofria, T. Collier, (2000). Evolution of biological complexity. PNAS, 97:4463-4468.

Adams F., (2003), The informational turn in philosophy, *Minds and Machines*, 13:471-501.

Albert R. and A. Barabasi, (2002). Statistical mechanics of complex networks. Rev. Mod. Phys. 74:47-97.

Alred D, (1998). Antigenic variation in Babesia Bovis: how similar is it to that of Plasmodium Falciparum? Ann. Trop. Med. Parasit. 92:461-472.

Atlan H. and I. R. Cohen, (1998), Immune information, self-organization and meaning, International Immunology 10:711-717.

Binney J., N. Dowrick, A. Fisher, and M. Newman, (1986), The theory of critical phenomena. Clarendon Press, Oxford UK.

Bonner J., (1980), *The Evolution of Culture in Animals*, Princeton University Press, Princeton, NJ.

Cohen I.R., (1992), The cognitive principle challenges clonal selection. Immunology Today 13:441-444.

Cohen I.R., (2000), *Tending Adam's Garden: evolving the cognitive immune self*, Academic Press, New York.

Combes C., (2000), Selective pressure in host-parasite systems [in French]. J. Soc. Biol. 194:19-23.

Cover T. and J. Thomas, (1991), *Elements of Information Theory*. Wiley, New York.

Dimitrov A. and J.Miller, (2001), Neural coding and decoding: communication channels and quantization. Network: Comput. Neural Syst. 12:441-472.

DiNola J., and M. Neuberger, (2002), Altering the pathway of immunoglobin hypermutation by inhibiting uracil-DNA glycosylase, Nature, 419:43-48.

Dretske F., (1981), *Knowledge and the flow of information*, MIT Press, Cambridge, MA.

Dretske F., (1988), *Explaining behavior*, MIT Press, Cambridge, MA.

Dretske F., (1992), What isn't wrong with folk psychology, *Metaphilosophy*, 29:1-13.

Dretske F., (1993), Mental events as structuring causes of behavior. In *Mental causation*, (ed. A. Mele and J. Heil), pp. 121-136, Oxford University Press.

Dretske F., (1994), The explanatory role of information, *Philosophical Transactions of the Royal Society A*, 349:59-70.

Eldredge N., (1985), *Time Frames: The Rethinking of Darwinian Evolution and the Theory of Punctuated Equilibria*, Simon and Schuster, New York.

Ewald P., (2000), *Plague Time: How Stealth Infections Cause Cancers, Heart Disease, and Other Deadly Ailments*, The Free Press, New York.

Gearhart P., (2002), The roots of antibody diversity. Nature 419:29-31.

Grossman Z., (1989), The concept of idiotypic network: deficient or premature? In: H. Atlan and I.R. Cohen (eds.), *Theories of Immune Networks*, Springer Verlag, Berlin, p. 3852.

Hartl D. and A. Clark, (1997). *Principles of Population Genetics*, Sinaur Associates, Sunderland MA.

Levins R. and R. Lewontin, (1985). *The Dialectical Biologist*, Harvard University Press, Cambridge MA.

Levins R, (1998), The internal and external in explanatory theories. Science as Culture. 7:557-582.

Lewontin R., (2000), *The Triple Helix: Gene, Organism and Environment.* Harvard University Press, Cambridge MA.

McCauley L., (1993), *Chaos, Dynamics, and Fractals: An Algorithmic Approach to Deterministic Chaos*, Cambridge University Press, UK.

Nisbett R., K. Peng, C. Incheol and A. Norenzayan, (2001). Culture and systems of thought: holistic vs. analytic cognition. Psychological Review 108: 291-310.

Nowak M. and R. May, (2000). *Virus dynamics: the mathematical foundations of immunology and virology.* Oxford University Press, New York.

Oyama S., P.E. Griffiths, and R.D. Gray, (2001), *Cycles of Contingency: Developmental Systems and Evolution*, The MIT Press, Cambridge, MA.

Pielou E, (1977), *Mathematical Ecology*. John Wiley and Sons, New York.

Richerson P. and R. Boyd, (1995), The evolution of human hypersociality. Paper for Ringberg Castle Symposium on Ideology, Warfare and Indoctrinability (January, 1995), and for HBES meeting.

Ridley M., (1996), *Evolution*, Second Edition, Blackwell Science, Oxford, UK.

Rojdestvenski I. and M. Cottam, (2000), Mapping of statistical physics to information theory with applications to biological systems. J. Theor. Biol. 202: 43-54.

Schoepf B.G., C. Schoepf, and J.V. Millen, (2000), Theoretical Therapies, Remote Remedies: SAPs and the Political Ecology of Poverty and Health in Africa. In: J.Y. Kim, J.V. Millen, A. Irwin, and J Gershman (eds.), *Dying for Growth: Global Inequality and the Health of the Poor*, Common Courage Press, Monroe, ME, p. 91.

Shirkov D. and V. Kovalev, (2001), The Bogoliubov renormalization group and solution symmetry in mathematical physics. Phys. Reports 352: 219-249.

Tauber A., (1998), Conceptual shifts in immunolog: Comments on the 'two-way paradigm'. In K Schaffner and T Starzl (eds.), Paradigm changes in organ transplantation, *Theoretical Medicine and Bioethics*, 19:457-473.

Tenallion O., F. Taddei, M. Radman, and I. Matic, (2001), Second order selection in bacterial evolution: selection acting on mutation and recombination rates in the course of adaptation, Research in Microbiology, 152:11-16.

Thaler D., (1999), Hereditary stability and variation in evolution and development, Evolution and Development, 1:113-122.

Villarreal L.P., V.R. Defilippis, and K.A. Gottlieb, (2000), Acute and persistent viral life strategies and their relationship to emerging diseases, *Virology*, 272:1-6.

Wallace D. and R. Wallace., (1998), *A Plague on Your Houses: How New York Was Burned Down and National Public Health Crumbled*, Verso, New York.

Wallace R., (2000), Language and coherent neural amplification in hierarchical systems: Renormalization and the dual information source of a generalized spatiotemporal stochastic resonance, Int. J. Bifurcation and Chaos 10:493-502.

Wallace R., (2002a), Immune cognition and vaccine strategy: pathogenic challenge and ecological resilience, Open Sys. Inform. Dyn. 9:51-83.

Wallace R., (2002b), Adaptation, punctuation and rate distortion: non-cognitive 'learning plateaus' in evolutionary process, Acta Biotheoretica, 50:101-116.

Wallace R. and D. Wallace, (2003), Structured psychosocial stress and therapeutic failure. http://www.arxiv.org/abs/q-bio.NC/0310005.

Wallace R. and R.G. Wallace, (1998), Information theory, scaling laws and the thermodynamics of evolution, Journal of Theoretical Biology 192:545-559.

Wallace R. and R.G. Wallace, (1999), Organisms, organizations and interactions: an information theory approach to biocultural evolution, BioSystems 51:101-119.

Wallace R. and R.G. Wallace, (2002), Immune cognition and vaccine strategy: beyond genomics, Microbes and Infection 4:521-527.

Wallace R., R.G. Wallace and D. Wallace, (2003), Toward cultural oncology: the evolutionary information dynamics of cancer, Open Sys. Inform. Dynam., 10:159-181.

Wallace R., D. Wallace and R.G. Wallace, (In press), Biological limits to reduction in rates of coronary heart disease: a punctuated equilibrium approach to immune cognition, chronic inflammation, and pathogenic social hierarchy, JNMA.

Wallace, R.G., (In press), Projecting the impact of HAART on the evolution of HIV's life history, Ecological Modelling.

Wilson K.,(1971), Renormalization group and critical phenomena. I Renormalization group and the Kadanoff scaling picture. Phys. Rev. B. 4: 3174-3183.